\documentclass[aps,prd,twocolumn,groupedaddress,eqsecnum,floatfix]{revtex4}
\input epsf

\begin{document}
\preprint{MADPH-03-1324}

\title{Reduction of the QCD string to a time component vector potential} 
\author{Theodore J. Allen}
\affiliation{Physics Department, Hobart \& William Smith Colleges \\
Geneva, New York 14456 USA}

\author{M. G. Olsson}
\affiliation{Department of Physics, University of Wisconsin, \\
1150 University Avenue, Madison, Wisconsin 53706 USA }

\date{\today}

\begin{abstract}
We demonstrate the equivalence of the relativistic flux tube model of
mesons to a simple potential model in the regime of large radial
excitation.  We make no restriction on the quark masses; either quark may
have a zero or finite mass.  Our primary result shows that for fixed
angular momentum and large radial excitation, the flux tube/QCD string
meson with a short-range Coulomb interaction is described by a spinless
Salpeter equation with a time component vector potential $V(r) = ar -
k/r$.
\end{abstract}
\maketitle

\section{Introduction}\label{sec:intro}

Although Quantum Chromodynamics (QCD) is almost surely the correct theory
of strong interactions, it remains difficult to explore its predictions in
the non-perturbative regime.  For hadron states the non-perturbative, or
confinement, regime corresponds to large distances.  For mesons confinement
dominates the dynamics of even the heaviest quark states.  It has long been
suspected that when the color sources are widely separated, the color
electric fields collapse into relatively thin configurations known as flux
tubes, or QCD strings.  The evidence for such string-like configurations is
primarily:
\begin{itemize}
\item universal linear Regge trajectories, reflecting a linear confining %
potential and relativistic kinematics \cite{ref:one};
\item lattice simulation of the energy density \cite{ref:two};
\item relativistic corrections of the flux tube model \cite{ref:three}
agree with those of Wilson loop QCD \cite{ref:nine};
\item for heavy onia, the flux tube model reduces to the very successful
linear confinement potential model;
\item agreement of the vibrating string picture \cite{ref:four} with lattice %
simulations of excited QCD states with fixed sources \cite{ref:ten}.
\end{itemize}

The relativistic string/quark model for mesons can be exactly solved
numerically for arbitrary quark masses \cite{ref:six}.

An excellent approximation for the boundstate energy $E$ of a meson
consisting of one massive and one massless quark is the spectroscopic
relation 
\begin{equation}\label{eq:one}
{E^2\over \pi a} = L + 2n + \frac32 .
\end{equation}
Here the angular momentum quantum number $L$ and radial quantum number $n$
take values $0,1,2,\ldots$.  The constant $a$ is the tension of the QCD
string.  For a meson with two massless quarks, the spectroscopic
relationship becomes \cite{ref:five}
\begin{equation}\label{eq:two}
{E^2\over 2 \pi a} = L + 2n + \frac32 .
\end{equation}
Recently we have shown semi-classically that in the limit of large radial
excitation, Eq.~(\ref{eq:one}) and Eq.~(\ref{eq:two}) follow in both the flux
tube model and in linear time component vector
confinement \cite{ref:five}. 

In this paper we demonstrate that the wave equation for the flux tube meson
with a short-range Coulomb interaction can be reduced to a spinless
Salpeter equation with a pure time component vector (TCV) potential
interaction.  This reduction holds rigorously for large radial excitation
but is accurate for all states.  This result establishes the close
connection between flux tube dynamics and the TCV potential model for
spinless quarks.

In section \ref{sec:string} we discuss the dynamics of a flux tube with
quarks of arbitrary masses.  The critical approximation that the string can
be assumed to be non-relativistic is shown to be accurate for $n\gg L$ in
section \ref{sec:three}.  In section \ref{sec:four} we establish the TCV
spinless Salpeter equation from the QCD string.  We conclude in section
\ref{sec:five}.  We include the detailed algebraic steps for sections
\ref{sec:three} and \ref{sec:four} in the Appendix.

\section{The QCD string with arbitrary quark masses}\label{sec:string}

There are two equivalent methods of extracting the conserved quantities of
the spinless quark-string system.  The momentum-energy approach considers a
straight color electric tube of energy $a$ per unit length.  From Lorentz
boosting a string element perpendicular to its orientation, the momentum,
angular momentum, and energy of the string are easily
obtained \cite{ref:three,ref:six}.  This intuitive construction is appealing
for its simplicity.

One can also extract the conserved quantities more formally by using
Noether's theorem and an action formalism.  We take an action
\cite{ref:seven} consisting of two pieces: one piece, the Nambu-Goto
action, is proportional to the invariant surface area swept out by the
string connecting the two quarks, the other piece is a sum of two terms,
each proportional to the invariant length of a quark worldline:
\begin{eqnarray}
S & = &-m_1\int d\tau \sqrt{-\dot{x}_1^2}\,
 - m_2\int d\tau \sqrt{-\dot{x}_2^2} + \nonumber \\
&& -{a} \int d\tau\int_{\sigma_1}^{\sigma_2} d\sigma\, \sqrt{(\dot{X}\cdot
X^\prime)^2 - (\dot X)^2 (X^\prime)^2}, \nonumber  \\
& = & \int d\tau \left(L_q + \int_{\sigma_1}^{\sigma_2} d\sigma \,
{\cal L}_s\right) .
\label{eq:action}
\end{eqnarray}
In this expression $x^\mu_1$ and $x^\mu_2$ are the positions of the
quarks, and $X^\mu(\sigma,\tau)$ is the position of the string.  The quark
masses are $m_1$ and $m_2$, $a$ is the string tension, a dot denotes a
derivative with respect to $\tau$, and a prime denotes a derivative with
respect to $\sigma$.  In our notation, the dot product between two four
vectors $A$ and $B$, uses a metric of signature $+2$; $A\cdot B = A^\mu
B^\nu \eta_{\mu\nu} = - A^0 B^0 + {\bf A}\cdot{\bf B}$.  The string
terminates on a quark at each end
\begin{equation}\label{eq:endpoints}
X^\mu (\sigma_1) = x_2^\mu, \quad X^\mu(\sigma_2) = x_1^\mu . 
\end{equation}

The momentum and angular momentum of the quark-string system can be found
from Noether's theorem.  The action is invariant under the combination of a
boost and a translation
\begin{eqnarray}\label{eq:poincare}
\delta x^\mu_i & = & a^\mu + \omega^{\mu}_{\phantom{\mu}\nu} x_{i}^{\nu} 
, \nonumber \\
\delta X^\mu(\sigma) & = & a^\mu + \omega^{\mu}_{\phantom{\mu}\nu}
X^\nu(\sigma) . 
\end{eqnarray}
The variation of the finite time action under (\ref{eq:poincare}) is the
sum of the translation times momentum change and rotation times angular
momentum change. 
\begin{eqnarray}
0 & = & \delta\int_{\tau_1}^{\tau_2} d\tau \left(L_q + \int d\sigma\,{\cal
L}_s\right)  \\  
& = & a^\mu\left[P_\mu(\tau_2) - P_\mu(\tau_1)\right] +
\frac12\omega^{\mu\nu} \left[ L_{\mu\nu}(\tau_2) -
L_{\mu\nu}(\tau_1)\right] . \nonumber
\end{eqnarray}
The resulting momentum and angular momentum are the canonical ones,
\begin{eqnarray}\label{eq:P}
P_\mu & = & {\partial L_q\over \partial \dot{x}_1^\mu } +
{\partial L_q\over \partial \dot{x}_2^\mu } +
\int_{\sigma_1}^{\sigma_2} d\sigma\, {\partial{\cal L}_s\over\partial
\dot{X}^\mu}  \nonumber \\
& = & m_1 {\dot{x}_{1\mu} \over \sqrt{-\dot{x}_1^2} } + m_2 {\dot{x}_{2\mu}
\over \sqrt{-\dot{x}_2^2} } + \\
&\phantom{=} & +\> a \int_{\sigma_1}^{\sigma_2}d\sigma\,  {\dot{X}_\mu
(X^\prime)^2 - X_\mu^\prime (\dot{X}\cdot X^\prime) \over \sqrt{(\dot{X}\cdot
X^\prime)^2 - (\dot X)^2 (X^\prime)^2}}\> , \nonumber
\end{eqnarray}

\begin{eqnarray}\label{eq:J}
L_{\mu\nu} & = & x_{1[\mu}{\partial L_q\over \partial \dot{x}_1^{\nu]} } +
x_{2[\mu} {\partial L_q\over \partial \dot{x}_2^{\nu]} } +
\int_{\sigma_1}^{\sigma_2} d\sigma\, X_{[\mu}(\sigma){\partial{\cal
L}_s\over\partial \dot{X}^{\nu]}} \nonumber \\
& = & m_1 {x_{1[\mu}\dot{x}_{1\nu]} \over \sqrt{-\dot{x}_1^2} } + m_2
{x_{2[\mu}\dot{x}_{2\nu]} \over \sqrt{-\dot{x}_2^2} } + \\
& + & a \int_{\sigma_1}^{\sigma_2} \kern-3pt d\sigma \,
{X_{[\mu}\dot{X}_{\nu]} (X^\prime)^2 - X_{[\mu} X_{\nu]}^\prime
(\dot{X}\cdot X^\prime) \over \sqrt{(\dot{X}\cdot X^\prime)^2 - (\dot X)^2
(X^\prime)^2}}\> . \nonumber
\end{eqnarray}

The action (\ref{eq:action}), momentum (\ref{eq:P}), and angular momentum
(\ref{eq:J}) are invariant under changes of variables in $\sigma$ and
$\tau$,
\begin{equation}\label{eq:diffeo}
(\sigma,\tau) \mapsto \left(\tilde\sigma(\sigma,\tau),
\tilde\tau(\sigma,\tau)\right) ,
\end{equation}
as long as the boundaries remain at values of $\tilde\sigma$ that are
independent of $\tilde\tau$.

We use this coordinate invariance to choose the parameter $\tau$ to be
the laboratory time, and the endpoints to be at particular values of
$\sigma_1$ and $\sigma_2$;
\begin{eqnarray}\label{eq:timegauge}
x_1^0 & = & x_2^0 = X^0(\sigma,\tau) = \tau = t , \\
\sigma_1 & = & 0, \quad \sigma_2 = 1 .
\end{eqnarray}
The straight string approximation to the equations of motion is exact for
uniform circular motion of the quarks and is a good approximation in
realistic mesons \cite{ref:eight}.  In this approximation, the
position of the string at any time $\tau$ lies along a straight line
between the quarks, which we can parameterize linearly as
\begin{equation}\label{eq:straight}
{\bf X}(\sigma)  =  (1-\sigma){\bf x}_2 + \sigma {\bf x}_1 .
\end{equation}
We further denote the separation between the quarks at a particular instant
as
\begin{equation}\label{eq:R}
{\bf r} \equiv {\bf x}_1 - {\bf x}_2 .
\end{equation}

Under the approximation (\ref{eq:straight}), the energy of the system
becomes
\begin{eqnarray}\label{eq:P0}
P^0 & = & {m_1\over \sqrt{1 - {v}_1^2}} + {m_2\over \sqrt{1 - {
v}_2^2}} + \nonumber \\
&& + \> a \int_0^1 d\sigma\, {r^2 \over \sqrt{ ({\bf r}\cdot {\bf
v}(\sigma))^2 - r^2(-1 + {v}^2(\sigma))}}\nonumber \\
& = & m_1\gamma_1 + m_2\gamma_2 + ar \int_0^1 d\sigma\,{1\over \sqrt{1 -
{v}_\perp^2(\sigma)}} ,
\end{eqnarray}
where ${\bf v}_1 = d{\bf x}_1/dt$ and ${\bf v}_2 = d{\bf x}_2/dt$ are the
quark velocities and we have defined 
\begin{eqnarray}
{\bf v}(\sigma) & = & \dot{\bf X}(\sigma) = (1-\sigma){\bf v}_2 + \sigma
{\bf v}_1 , \nonumber \\
{\bf v}_\perp(\sigma)  & = & {\bf v}(\sigma) - {{\bf r}\cdot{\bf v}(\sigma)
\over r}\, {{\bf r}\over r} .
\end{eqnarray}
The momentum of the system with the straight string approximation becomes
\begin{eqnarray}\label{eq:3P}
{\bf P} & = & {m_1 {\bf v}_1\over \sqrt{1 - v_1^2}} + {m_2 {\bf v}_2\over
\sqrt{1 - v_2^2}} + \\ 
&& +\> a \int_0^1 d\sigma\, {{\bf v}(\sigma) r^2 - {\bf
r}({\bf v}(\sigma)\cdot{\bf r}) \over \sqrt{ ({\bf r}\cdot {\bf
v}(\sigma))^2 - r^2(-1 + {v}^2(\sigma))}} \nonumber \\
& = & m_1\gamma_1{\bf v}_1 + m_2\gamma_2{\bf v}_2 + ar\int_0^1d\sigma\, {{\bf
v}_\perp(\sigma) \over \sqrt{1 - {v}_\perp^2(\sigma)}} . \nonumber
\end{eqnarray}
The vector angular momentum is obtained from $L_{\mu\nu}$ as $L^i =
\frac12\epsilon^{ijk} L_{jk}$, which, for a straight string, becomes
\begin{eqnarray}\label{eq:Jz}
{\bf L} & = & m_1{{\bf x}_1\times {\bf v}_1\over\sqrt{1-v_1^2}} + m_2 {{\bf
x}_2\times {\bf v}_2 \over\sqrt{1-v_1^2}} + \nonumber \\ 
&& +\> a\int_0^1d\sigma\,{r^2{\bf X}(\sigma)\times{\bf v}(\sigma) - {\bf
X}(\sigma)\times {\bf r}({\bf r}\cdot{\bf v}(\sigma)) \over\sqrt{ ({\bf
r}\cdot {\bf v}(\sigma))^2 - r^2(-1 + {v}^2(\sigma))}} \nonumber \\
&=& m_1\gamma_1\, {\bf x}_1\times {\bf v}_1 + m_2\gamma_2\, {\bf x}_2\times
{\bf v}_2 + \nonumber \\
&& +\> ar \int_0^1d\sigma\,{{\bf X}(\sigma)\times{\bf v}_\perp(\sigma)
\over \sqrt{1 - {v}_\perp^2(\sigma)}} .
\end{eqnarray}

We evaluate the integrals in Eqs.~(\ref{eq:P0}), (\ref{eq:3P}), and
(\ref{eq:Jz}) for a string rotating about the $z$-axis and instantaneously
lying along the $x$-axis from $-r_2$ to $r_1$.  The string's perpendicular
velocity runs from $-v_{\perp 2}$ to $v_{\perp 1}$;
\begin{eqnarray}
X(\sigma) & = & \sigma r_1 + (\sigma -1) r_2 , \nonumber \\
v_\perp(\sigma) & = & \sigma v_{\perp 1} + (\sigma - 1) v_{\perp 2} .
\end{eqnarray}
The total length of the string is $r = r_1 + r_2$.  Because we assume that
the string stays straight as it rotates, we have
\begin{equation}\label{eq:rot}
{v_{\perp 1}\over r_1 } = {v_{\perp 2} \over r_2} .
\end{equation}
Equation~(\ref{eq:rot}) implies
\begin{equation}\label{eq:ratio}
{r\over v_{\perp 1} + v_{\perp 2}} = {r_1 \over v_{\perp 1} } = {r_2 \over v_{\perp 2}} .
\end{equation}
We denote the non-quark pieces of the energy, momentum, and angular
momentum by $\lambda_M$, $\lambda_P$, and $\lambda_L$ respectively.  The
only contribution to $\lambda_P$ and $\lambda_L$ is from the string.  The
energy $\lambda_M$ has contributions from the Coulomb piece as well as the
string.  We find the string energy in Eq.~(\ref{eq:P0}) to be
\begin{eqnarray}\label{eq:P0string}
ar \int_0^1 d\sigma\, {1\over \sqrt{1 -v_\perp^2}} & = & {ar
\left(\sqrt{1-v_{\perp2}^2} - \sqrt{1 - v_{\perp 1}^2} \right) \over
v_{\perp 1} + v_{\perp 2}} \nonumber \\
& = & {ar \over v_{\perp 1}
+ v_{\perp 2}}(\gamma_{\perp 2}^{-1} - \gamma_{\perp 1}^{-1} ) \\
& \equiv & \lambda_M + {k\over r} \nonumber . 
\end{eqnarray}
The perpendicular momentum of the string from Eq.~(\ref{eq:3P}) is
\begin{eqnarray}\label{eq:Pstring}
ar \int_0^1 d\sigma\, {v_\perp \over \sqrt{1 -v_\perp^2}} & = & {ar \left(
\arcsin(v_{\perp 1}) + \arcsin(v_{\perp 2}) \right) \over v_{\perp 1} +
v_{\perp 2}} \quad \nonumber \\
& \equiv& \lambda_P . 
\end{eqnarray}
The angular momentum of the string from Eq.~(\ref{eq:Jz}) becomes
\begin{eqnarray}\label{eq:Jstring}
 ar \int_0^1d\sigma\,{X\, v_\perp \over \sqrt{1 -
v_\perp^2}}  =   {ar \over 2( v_{\perp 1}
+ v_{\perp 2})} \Big[ - r_1 \gamma_{\perp 1}^{-1} + \nonumber  \\ 
 - \> r_2 \gamma_{\perp 2}^{-1} +  
 {ar \over v_{\perp 1} + v_{\perp 2}} \big[\arcsin(v_{\perp 1}) +
\arcsin(v_{\perp 2})\big] \nonumber \\
+\> (\gamma_{\perp 2}^{-1} - \gamma_{\perp 1}^{-1}) {v_{\perp 2} r_1 -
v_{\perp 1} r_2 \over v_{\perp 1}
+ v_{\perp 2}} \Big] \equiv \lambda_L . \qquad
\end{eqnarray}

We can simplify the quark pieces of the conserved quantities through use
of the identity
\begin{eqnarray}
{p_r^2 + m^2 \over m^2 } & = & {\gamma^2 v_r^2 + 1} = {v_r^2 \over 1 -
v_r^2 - v_\perp^2} + 1 \nonumber \\
& = & {1 - v_\perp^2 \over 1 - v_r^2 - v_\perp^2} 
\equiv {\gamma^2 \over \gamma_\perp^2},
\end{eqnarray}
or
\begin{equation}\label{eq:Wgamma}
m\gamma = \sqrt{p_r^2 + m^2} \, \gamma_\perp \equiv W_r \gamma_\perp .
\end{equation}

Using Eqs.~(\ref{eq:P0string}), (\ref{eq:Pstring}), (\ref{eq:Jstring}), and
(\ref{eq:Wgamma}), we obtain our final expressions for the conserved
quantities
\begin{eqnarray}
P^0 & = & W_{r 1}\gamma_{\perp 1} + W_{r 2}\gamma_{\perp 2} + \\
&& +\> ar\left({r_1\over v_{\perp 1}} \arcsin(v_{\perp 1}) + {r_2\over
v_{\perp 2}} \arcsin(v_{\perp 2})\right) , \quad \nonumber \\
P_\perp & = & W_{r 1}\gamma_{\perp 1}v_{\perp 1} -  W_{r 2}\gamma_{\perp
2}v_{\perp 2}  + \nonumber \\
&& +\> a\left( {r_2\over v_{\perp 2}}\gamma_{\perp 2}^{-1} - {r_1\over v_{\perp
1}}\gamma_{\perp 1}^{-1}\right) , \\
L & = & W_{r 1}\gamma_{\perp 1}v_{\perp 1}r_1 +  W_{r 2}\gamma_{\perp
2}v_{\perp 2}r_2 + \nonumber \\
&& + \> {a\over 2} {r_1^2\over v_{\perp 1}}\left({\arcsin(v_{\perp 1})\over
v_{\perp 1}} - \gamma_{\perp 1}^{-1}\right) + \nonumber \\
&& + \> {a\over 2} {r_2^2\over v_{\perp 2}}\left({\arcsin(v_{\perp 2})\over
v_{\perp 2}} - \gamma_{\perp 2}^{-1}\right) .
\end{eqnarray}
These results are exactly the starting point in the momentum-energy
approach \cite{ref:three,ref:six}.

Finally, we note that the Nambu-Goto Lagrangian for a straight string reduces
to 
\begin{eqnarray} \label{eq:StringS}
L_{\rm string} & = & -ar \int_0^1 d\sigma\, \sqrt{1 -
v_\perp^2(\sigma)}\nonumber  \\
&=& -\,{a\over 2} \Bigg({r_1\over v_{\perp 1}} \arcsin(v_{\perp 1}) +
{r_2\over v_{\perp 2}} \arcsin(v_{\perp 2})  \nonumber \\
&&  \quad +\> r_1 \gamma_{\perp 1}^{-1} +
r_2 \gamma_{\perp 2}^{-1}\Bigg) . 
\end{eqnarray}

\section{The high radial excitation regime}\label{sec:three}

We now make the crucial observation that, for fixed angular momentum and
large radial excitation, we may assume the string 
velocity is small without changing the meson dynamics.  This is because
$v_\perp$ only becomes large near the inner turning point, where it reaches
$v_\perp=1$ in the case of a massless quark. However, near the inner
turning point the string is short for very radial orbits so it carries
little angular momentum or energy.  Henceforth, for notational convenience,
we generally suppress the $\perp$ subscript, and let $v$ and $\gamma$
denote $v_\perp$ and $\gamma_\perp$.

To demonstrate our approximation, we consider the equal quark mass
case $m_1=m_2$, from which follows $v_1 = v_2 \equiv v$, and $r_1 = r_2 =
r/2$.  We define
\begin{eqnarray}
S(v) & = & {\arcsin (v) \over v} , \\
f(v) & = & {1\over 2v}\left( S - \sqrt{1-v^2}\right) , 
\end{eqnarray}
and hence Eqs.~(\ref{eq:P0string}), (\ref{eq:Pstring}) and
(\ref{eq:Jstring}) (or Eqs.~(\ref{eq:lambdaP}), (\ref{eq:lambdaL}), and
(\ref{eq:lambdaM})) can be expressed as
\begin{eqnarray}
\lambda_P & = & 0, \\
\lambda_L & = & \frac12 a r^2 f \\
\lambda_M & = & a r S - \frac{k}r .
\end{eqnarray}
The total energy $E$ and angular momentum $L$ of the system are related to
the quark separation and transverse velocities by Eq.~(\ref{eq:a8}).  In
this notation the relation (\ref{eq:a8}) becomes
\begin{equation}\label{eq:34}
v\left( E - arS + {k\over r}\right) + arf = {2L\over r}.
\end{equation}

To reach reasonably general numerical conclusions we define the
dimensionless quantities
\begin{equation}
x = {ar\over E}, \quad \beta = {aL\over E^2}, \quad \kappa =
{ak\over E^2} .
\end{equation}
In these dimensionless variables, Eq.~(\ref{eq:34}) becomes
\begin{equation}\label{eq:3.6}
x v \left(1 - xS + {\kappa\over x}\right) + x^2 f = 2\beta .
\end{equation}

\begin{figure}[hbp]
\epsfxsize = \linewidth 
\hspace*{-5mm}
\epsfbox{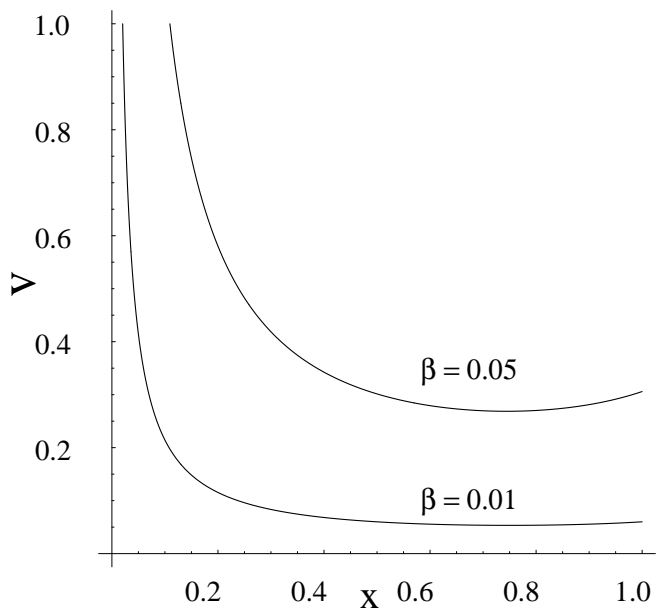}
\vskip 1 cm
\caption{Solution to Eq.~(\protect\ref{eq:3.6}) for the perpendicular
velocity of the string end as a function of distance $x=ar/E$ for two
values of $\beta = aL/E^2$ with no Coulomb potential; $\kappa = 0$.  For
any attractive Coulomb interaction ($\kappa > 0$) the velocities are lower,
making the approximation $v \ll 1$ better.  We observe that for radially
excited states $\beta \ll 1$ and $v\rightarrow 0$ except at the inner
turning point.}
\label{fig:one}
\end{figure}

For fixed $L$ and increasing $E$, we expect both $\beta$ and $\kappa$
to be small and $xv \approx 2\beta$.  We thus expect $v$ to be small
unless $x$ is small.  In Fig.~\ref{fig:one} we show the exact numerical
solution of Eq.~(\ref{eq:3.6}) for two values of $\beta$.  For simplicity we
consider no short range interaction ($\kappa = 0$.)

\begin{figure}[hbp]
\epsfxsize = \linewidth 
\hspace*{-5mm}
\epsfbox{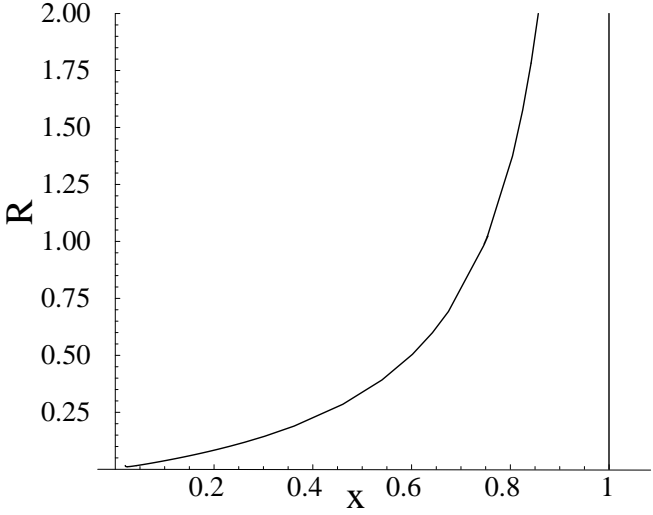}
\vskip 1 cm
\caption{The ratio $R$ of string to quark angular momentum as a function of
dimensionless distance $x$ as in Eq.~(\protect\ref{eq:dimless}) with $\beta
= 0.01$.  The string angular momentum is negligible for small quark
separation.}
\label{fig:two}
\end{figure}

It remains to demonstrate that the ratio of string to quark angular
momentum is only appreciable when $v$ is small.  From the total angular
momentum relation (\ref{eq:a2}) we find
\begin{equation}\label{eq:3.8}
R \equiv {\rm string\ angular\ momentum \over quark\ angular\ momentum} =
{\lambda_L\over 2\Omega v{r\over2}} ,
\end{equation}
where $\Omega \equiv W_r \gamma_\perp$ is the quark kinetic energy.
The total system energy, Eq.~(\ref{eq:a3}), becomes
\begin{equation}\label{eq:311}
E = 2\Omega + arS - {k\over r} .
\end{equation}
In terms of dimensionless parameters, again in the case $k = 0 =
\kappa$, by using Eq.~(\ref{eq:311}) we have
\begin{equation}\label{eq:dimless}
R = {x f(v)\over v (1 - xS)}.
\end{equation}
We show $R$ as a function of $x$ in Fig.~\ref{fig:two} for $\beta=0.01$ by
using Eq.~(\ref{eq:3.6}).  As we stated at the beginning of this section,
the string angular momentum becomes small compared to either quark's near
the inner turning point.  This tells us that, in the radially excited
regime, we are justified in assuming that $v_\perp$ is small for the
string.  In the regions in which this is not true, the string contributes
negligibly to the energy and angular momentum.  Our approach will be to
keep full relativistic kinematics for the quarks and to assume that the
string velocity is small.

\section{Reduction to the spinless Salpeter wave equation}
\label{sec:four}

In the preceding section we examined the exact numerical solution of the
QCD string equations for the equal mass case, $m_1=m_2$.  We saw that if
$aL\ll E^2$ the string perpendicular velocity can be assumed to be small
without changing any dynamical result.  In the appendix we examine the
detailed algebraic consequences of this approximation.  We begin this
section by outlining the main results found there.  

In section \ref{sec:string} we established the total perpendicular
momentum, angular momentum, and energy for arbitrary masses $m_1$ and $m_2$
at the ends of a straight QCD string.  The results are expressed in terms
of the quark perpendicular velocities and distances from the center of
momentum point.  This point is defined by the condition $P_\perp = 0$.

Our strategy will be first to set up the relations defining $v_1 \equiv
v_{1\perp}$ and $v_2 \equiv v_{2\perp}$ in terms of $r=r_1 + r_2$, the
state mass $M$, and the quark masses.  We obtain an explicit solution for
$v_1$ and $v_2$ in the desired limit of small velocities.  The result
confirms  the numerical result of the preceding section in that $v_i
\rightarrow 0$ for large excitation energy $E$ assuming the angular momentum
$L$ is fixed.  It then becomes evident that in this limit the string does not
enter the dynamics except for its static energy.  The TCV wave equation for
arbitrary masses then follows.  

Referring to the appendix, we use Eqs.~(\ref{eq:aO}) to (\ref{eq:a6})
to obtain the radial energy factors, Eqs.~(\ref{eq:a6.5}) and
(\ref{eq:a7}),
\begin{equation}\label{eq:vO}
v_i \Omega_i = {L\over r} + \alpha_i, \quad i=1,2,
\end{equation}
where $\Omega_i \equiv W_{ri} \gamma_{\perp i}$. Most of the remainder of
the appendix is devoted to finding the quantities $\alpha_1$ and
$\alpha_2$.

We use Eq.~(\ref{eq:vO}) to eliminate the $\Omega_i$ in the energy equation
(\ref{eq:a3}) and obtain Eq.~(\ref{eq:a8}),
\begin{equation}\label{eq:5.2}
M-\lambda_M = {L\over r} \left( \frac1{v_1} + \frac1{v_2} \right) -
{\lambda_L\over r} \left( \frac1{v_1} + \frac1{v_2} \right) + {\lambda_P}
\left( \frac1{v_2} - \frac1{v_1} \right) .
\end{equation}
This is our first relation between $v_1$ and $v_2$.  Next, we use the
definition of $\Omega_i = W_{ri}\gamma_{\perp\, i}$ given in
Eq.~(\ref{eq:a9}) to eliminate $p_r$ and obtain our second relation between
$v_1$ and $v_2$, Eq.~(\ref{eq:a10}); 
\begin{eqnarray}\label{eq:5.3}
&& {(M-\lambda_M)\over v_1v_2} \left[ {L\over r} - {\lambda_L\over r} +
{\lambda_P \left( v_1^2 + v_2^2 \right) \over v_1^2 - v_2^2 } \right] 
\nonumber \\
&& \quad -\, {\lambda_P\over v_1 - v_2} \left[ {2L\over r } - {2\lambda_L
\over r} + \lambda_P{v_1 - v_2\over v_1 + v_2} \right] \\
&& = {m_2^2 - m_1^2 \over v_1 -v_2} . \nonumber
\end{eqnarray}

By eliminating $L/r$ between the two relations (\ref{eq:5.2}) and
(\ref{eq:5.3}), we obtain the very useful relation given in (\ref{eq:a11});
\begin{eqnarray}\label{eq:5.4}
&& \left(M-\lambda_M\right)^2 + {2\lambda_P \over
v_1-v_2}\left(1-v_1v_2\right) \left(M - \lambda_M\right) + \lambda_P^2
\nonumber \\
&& = \left({m_2^2 - m_1^2 \over v_1 - v_2}\right) \left(v_1 + v_2\right) .
\end{eqnarray}
Up to this point, we have made no approximations concerning the velocities
$v_i$.  In the auxiliary relation (\ref{eq:5.4}) the terms involving
$v_1v_2$ and $\lambda_P^2$ are higher order in quark velocities.  We now
make the approximation of small quark velocities, so we drop these terms
and use the approximations of Eq.~(\ref{eq:a12}) to set $\lambda_P/(v_1 -
v_2) \simeq ar/2$. In the small quark velocity limit, the string energy,
$\lambda_M$, has no dependence on quark velocities, so we may approximate
Eq.~(\ref{eq:5.4}) as an equation for $(v_1 + v_2)/(v_1 - v_2)$ alone,
\begin{equation}\label{eq:5.5}
\left(m_2^2 - m_1^2\right)  \left({v_1 + v_2\over v_1 - v_2}\right) =
\left(M-\lambda_M\right) \left( M-\lambda_M + ar \right) ,
\end{equation}
which is Eq.~(\ref{eq:a13}).  From Eq.~(\ref{eq:5.5}) we
can find the ratio of the quark velocities;
\begin{eqnarray}
{v_2\over v_1} & = & {A-B\over A+B}, \label{eq:5.6} \\
A & = & \left( M - \lambda_M \right) \left( M - \lambda_M + ar\right)
,\label{eq:5.7} \\
B & = & \left( m_2^2 - m_1^2\right) .
\end{eqnarray}
We now use the small $v_i$ approximation of Eq.~(\ref{eq:a12}) to rework
Eq.~(\ref{eq:5.3}) as 
\begin{equation}\label{eq:5.8}
{m_2^2 - m_1^2 \over M - \lambda_M} \> {v_1 v_2\over v_1 - v_2} = {L\over
r} + {ar\over 6}\left(v_1 +  v_2 \right).
\end{equation}
Finally, we use Eq.~(\ref{eq:5.6}) for $v_2/v_1$ to obtain $v_1$ as a
function of $r$, 
\begin{equation}\label{eq:5.9}
{2L\over rv_1} = {A - B \over M - \lambda_M} - {2ar\over3}{ A \over (A+B) } .
\end{equation}
A similar expression obtains for $v_2$ by replacing $B$ by $-B$ in Eq.~(\ref{eq:5.9}).

Equation (\ref{eq:5.9}) is the key to obtaining our final goal of reducing
the string equations to a TCV wave equation.  We can now evaluate $\alpha_1$
from Eq.~(\ref{eq:vO}).  From the expression for $\alpha_1$ given in
Eq.~(\ref{eq:a6.5}) and the small $v_i$ expansions of Eq.~(\ref{eq:a12}),
we find
\begin{equation}\label{eq:5.10}
\alpha_1 = -\, {ar\over 6}\left( 2v_1 - v_2\right).
\end{equation}
Elimination of $v_2$ through the use of Eq.~(\ref{eq:5.6}) yields
\begin{equation}\label{eq:5.11}
\alpha_1 = -\, {a r v_1\over 6} \left({A + 3B \over A + B }\right) .
\end{equation}
The corresponding $\alpha_2$ follows by replacing $v_1$ by $v_2$ and $B$ by
$-B$.

Finally, we use Eq.~(\ref{eq:5.9}) to eliminate $v_1$ and find
\begin{equation}\label{eq:5.12}
\alpha_1 = -\,{aL\over 3}\left({(A+3B)\over {A^2 -B^2\over(M-\lambda_M)} -\frac23 
ar A}\right) .
\end{equation}
The angular momentum relation (\ref{eq:vO}) can then be expressed as
\begin{eqnarray}\label{eq:5.13}
v_1\gamma_1 W_{r1} & = & {L\over r} \left( 1 - F_1 \right), \\
F_1 & = & {(A + 3B) ar/3 \over {A^2 -B^2\over(M-\lambda_M)} -\frac23 
ar A} . \nonumber
\end{eqnarray}

The corresponding quantity, $F_2$ is obtained from Eq.~(\ref{eq:5.13}) by
the replacement of $B$ by $-B$,
\begin{equation}\label{eq:5.16}
F_2  =  {(A - 3B) ar/3 \over {A^2 -B^2\over(M-\lambda_M)} -\frac23 
ar A} . 
\end{equation}

We examine the two important limits of a ``light-light'' and a
``heavy-light'' meson.
\subsection{Light-light limit}

In the light-light limit, we take $m_1 = m_2 = 0$, so that $B=0$ and $M=E$.
In this case we have
\begin{equation}
F^{LL}_1 = F^{LL}_2 = {ar/3\over E + {k\over r} - \frac23 ar }
\end{equation}

\subsection{Heavy-light limit}

The heavy-light limit has $m_1 = 0$ and $m_2 \rightarrow \infty$.  In this
case we take $M= m_2 + E$ and find
\begin{eqnarray}
F^{HL}_1 & = & {ar/3\over E + {k\over r} - \frac23 ar } , \\
F^{HL}_2 & = & - \, \frac12 F^{HL}_1 .
\end{eqnarray}
The function $F_1$ varies by at most a factor of 2 over the whole mass
range $0 < m_2 < \infty$, while $F_2$ varies at most by a factor of 2, but
can change sign over that region.

We can solve for $\gamma_1$ from Eq.~(\ref{eq:5.13}) using $v_1\gamma_1 =
\sqrt{\gamma_1^2 -1 }$,
\begin{equation}\label{eq:5.18}
\Omega_1 = W_{r1}\gamma_1 = \sqrt{W_{r1}^2 + {L^2\over r^2}(1-F_1)^2}.
\end{equation}
This result is added to the corresponding $\Omega_2$, using $W_{ri} =
\sqrt{p_r^2 + m_i^2}$, to yield the Hamiltonian
(\ref{eq:a3}) 
\begin{eqnarray}\label{eq:5.19}
H & = & \Omega_1 + \Omega_2 + \lambda_M \strut \nonumber \\
& = & \sqrt{p_{r}^2 + {L^2\over
r^2}(1-F_1)^2 + m_1^2 }\, + \nonumber \\
&& +\> \sqrt{p_{r}^2 + {L^2\over r^2}(1-F_2)^2
+ m_2^2 } \, + \nonumber \\
& \phantom{=} & +\> ar - {k\over r}.
\end{eqnarray}
The spinless Salpeter equation,
\begin{equation}\label{eq:5.20}
H \psi = M \psi ,
\end{equation}follows from Eq.~(\ref{eq:5.19}).

To complete the argument, we now observe that the angular momentum term
$L^2/r^2$ is large only near the inner turning point which in turn is small
for highly radially excited states.  Near this inner turning point, the
$F_i$ factors are small since they are proportional to $r$.  Where the
$F_i$ become larger, their effect is negligible since their factors
multiply $L^2/r^2$.  Thus, for highly excited radial states, dropping the
$F_i$ will not change any dynamical results.  The subsequent Hamiltonian
(\ref{eq:5.19}) then becomes
\begin{equation}\label{eq:5.22}
H = \sqrt{p^2 + m_1^2} + \sqrt{p^2 + m_2^2} + ar - {k \over r} ,
\end{equation}
which is the TCV Hamiltonian with a TCV potential
\begin{equation}
V(r) = ar - {k\over r} .
\end{equation}

We note that at every step we have imposed the center of momentum rest
condition and hence the Hamiltonian, Eq.~(\ref{eq:5.22}), satisfies all
conservation laws, even for arbitrary masses.

\section{Discussion and Summary}
\label{sec:five}

We have shown that a relativistic QCD string meson model, with a short
range Coulomb part, reduces to a time component vector (TCV) potential
model for large radial excitations.  Our result follows for spinless
quarks with any masses from zero to infinite.  The result is remarkable
since the relativistic string generally dominates dynamically, carrying
both angular momentum and rotational energy.  However, for fixed angular
momentum and eccentric orbits, where $E^2 \gg aL$, the quarks act as if they
were moving in a static TCV potential that is linear at large distances and
Coulombic at short ones.  

The coincidence of the two systems in the radially excited regime results
from a confluence of effects.  For large radial excitation, the radial
velocities dominate over the perpendicular velocity, except near the inner
turning point, where the motion is rotation and the transverse velocities
reach light velocity in the extreme limit of a massless quarks.  However,
for large radial excitation, the inner turning point occurs at small
radius, so the string carries little angular momentum.  Thus we may assume
that the angular velocity is small everywhere and approaches zero as the
radial excitation increases.

\begin{figure}[hbp]
\epsfxsize = \linewidth 
\hspace*{-5mm}
\epsfbox{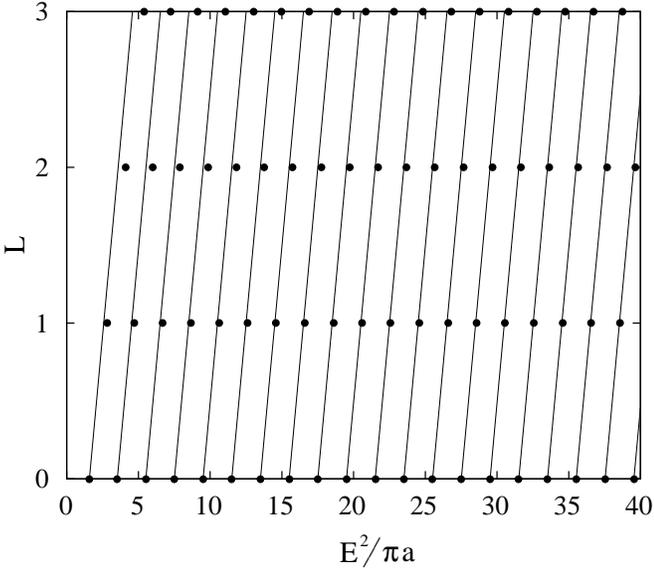}
\vskip 1 cm
\caption{The transition from a time component vector (TCV) confinement to
string dynamics.  The lines represent exact numerical solutions to the
string equations in the case $m_1 = 0$, $m_2=\infty$.  The dots are exact
numerical solutions to linear TCV confinement.  The squared excitation
energies ($E = M - m_2$) of the TCV system converge to those of the string
for large radial excitations with small angular momentum $L$.}
\label{fig:three}
\end{figure}

Conversely, the quark's radial energy is large and relativistic and
satisfies the spinless Salpeter equation given in Eqs.~(\ref{eq:5.20}) and
(\ref{eq:5.22}).   To show that the final result is correct, we observe in
Fig.~\ref{fig:three} two numerically exact sets of solutions.  The dots
represent the TCV solution with $k =0$ and a linear confinement
potential $V(r) = ar$.  The solid lines are interpolations of the QCD
string  solutions.  The latter are numerically exact quantized solutions of
the string equations (\ref{eq:aO}) to (\ref{eq:a6}).  For {\it s}-waves
($L=0$) there is no transverse motion ($v_\perp =0$) and the string equations
are exactly the {\it s}-wave TCV equation and the curves for each radial state
passes through the $L=0$ dots.  The non-trivial result of this paper is
seen for $L=1,2$, or $3$, where the curves come closer to the dots as the
radial excitation increases.

In previous work \cite{ref:five} we have approached this same result from a
different route.  If one quantizes the string semiclassically, one
reproduces the (numerically) exact results of the string spectroscopy for
highly radially excited states.  Conversely, if one semiclassically
quantizes the TCV spinless Salpeter equation with a linear confining
potential, again the string spectroscopy emerges for highly radially
excited states.  The work of this paper shows {directly} from the conserved
string quantities that TCV dynamics result for the highly radially excited
states.

Finally, the situation can be clarified by going back to the string action.
As we saw from Eq.~(\ref{eq:StringS}), the straight string Lagrangian can be
written as
\begin{equation}\label{eq:6.1}
L_{\rm string} = -ar \int_0^1 d\sigma\, \sqrt{1-v_\perp^2(\sigma)} .
\end{equation}
For small $v_\perp$, the string Lagrangian above expands to 
\begin{equation}\label{eq:6.2}
L_{\rm string} \simeq  -ar \left[1 + \frac16\left(v_{\perp 1}v_{\perp 2} - v_{\perp 1}^2 - v_{\perp 2}^2\right) + \cdots \right] .
\end{equation}

If one immediately sets $v_\perp \rightarrow 0$, the string action becomes
the linear piece of the TCV interaction Lagrangian.  One might worry
whether this is justifiable.  The present paper shows that it is.

In this work we began with the justification of the small $v_\perp$ string
approximation.  From the exact expressions for the straight string
momentum, angular momentum, and energy, we systematically approximate these
quantities for low string velocities.  We are then able to recast the
energy equation into the spinless Salpeter form and finally, to show that
this equation must be dynamically identical to the TCV equation in the
large radial excitation regime.

We might remind the reader that the strict constraint of a straight string
can also be relaxed.  Small deviations from straightness do not change the
conservation relations to first order and in ordinary hadrons the
deviations from straightness never become large \cite{ref:eight}.

\appendix
\section{}

In this appendix we give the detailed algebraic steps involved in writing
the energy of the quark/string system in terms of conserved quantities and
the quark separation.  We begin by finding the center of mass of the
system.  After several steps, we can express the quark velocities in terms
of the quark separation, the quark masses, and conserved quantities.

\subsection{Notation}

We use the straight string conserved quantities given in section
\ref{sec:string} and generally drop the $\perp$ subscripts on the quark
velocities and $\gamma\,$s for notational simplicity.  The coordinates $r_i$
and velocities $v_i$ are relative to the, as yet unknown, center of
momentum point.  The instantaneous positions of the quarks are taken to be
along the $x$-axis, with quark 1 at $x=r_1$ and quark 2 at $x=-r_2$.  Their
transverse velocities are $v_1$ and $-v_2$ respectively.  We also define
\begin{eqnarray}\label{eq:aO}
\Omega_i & = & W_{r\,i} \gamma_{\perp\, i}\,\strut , \\
W_{r\,i} & = & \sqrt{p_r^2 + m_i^2}\, , \\
\gamma_{\perp\, i} & = & {1\over \sqrt{ 1 - v_{\perp\, i}^2}}\, .
\end{eqnarray}

\subsection{Conserved quantities}
In section \ref{sec:string} we found the conserved quantities for the
quark and straight string system.  We gather these results together here.
The transverse momentum (\ref{eq:3P}) of the system is
\begin{equation}\label{eq:a1}
\Omega_1 v_{1} - \Omega_2 v_{2} + \lambda_P = 0,
\end{equation}
where 
\begin{equation}\label{eq:lambdaP}
\lambda_P =  {ar_2\over v_2} \gamma_2^{-1}  - {ar_1\over v_1} \gamma_1^{-1} .
\end{equation}

The angular momentum (\ref{eq:Jz}) consists of the contribution of the
quarks plus that of the string,
\begin{equation}\label{eq:a2}
L = \Omega_1 v_1 r_1 + \Omega_2 v_2 r_2 + \lambda_L ,
\end{equation}
with
\begin{equation}\label{eq:lambdaL}
\lambda_L = {a\over2} {r^2_1\over v_1} \left({\arcsin (v_1)\over v_1} -
\gamma_1^{-1}\right) +  {a\over2} {r^2_2\over v_2} \left({\arcsin
(v_2)\over v_2} - \gamma_2^{-1}\right) .
\end{equation}

The energy of the system (\ref{eq:P0}) is the sum of the quark energies and
the string energies
\begin{equation}\label{eq:a3}
M = \Omega_1 + \Omega_2 + \lambda_M ,
\end{equation}
where the string plus Coulomb contribution is
\begin{equation}\label{eq:lambdaM}
\lambda_M = a r_1 {\arcsin(v_1)\over v_1} + a r_2 {\arcsin(v_2)\over v_2} -
{k \over r} .
\end{equation}

\subsection{Straight string yields uniform angular velocity}

The straight string condition (\ref{eq:straight}) leads to equal quark
angular velocities;
\begin{equation}\label{eq:vr}
{v_1\over r_1 } = {v_2 \over r_2} .
\end{equation}
From Eq.~(\ref{eq:vr}) and the definition of the total quark separation
\begin{equation}
r = r_1 + r_2 ,
\end{equation}
we have
\begin{equation}\label{eq:a6}
r_i = r {v_i\over v_1+v_2} .
\end{equation}

\subsection{Quark kinetic energies}

The first step in finding the total energy of the system in terms of the
conserved quantities and the relative positions of the quarks is to write
the quark kinetic energies $\Omega_i$ in terms the system quantities $L$,
$r$, and the quark perpendicular velocities $v_i$.  We use
Eq.~(\ref{eq:a1}) and Eq.~(\ref{eq:a6}) to eliminate $\Omega_2$ in
(\ref{eq:a2}).  We find
\begin{eqnarray}\label{eq:a6.5}
v_1 \Omega_1 &= & {L\over r} + \alpha_1 , \\ 
\alpha_1 & = & - {\lambda_L \over r} - {v_2\over v_1 + v_2} \lambda_P ,
\nonumber
\end{eqnarray}
and
\begin{eqnarray}\label{eq:a7}
v_2 \Omega_2 &= & {L\over r} + \alpha_2 , \\
\alpha_2 & = & - {\lambda_L \over r} + {v_1\over v_1 + v_2} \lambda_P  . \nonumber
\end{eqnarray}

\subsection{Total energy}

Next we express the total energy (\ref{eq:a3}) in terms of the rewritten
$\Omega_i$'s above.  We find
\begin{equation}\label{eq:a8}
M-\lambda_M = {L\over r} \left( \frac1{v_1} + \frac1{v_2} \right) -
{\lambda_L\over r} \left( \frac1{v_1} + \frac1{v_2} \right) + {\lambda_P}
\left( \frac1{v_2} - \frac1{v_1} \right) .
\end{equation}

We can find another relation without $\Omega_i$'s by going back to the
definition of $\Omega_i$ and $W_{r\, i}$
\begin{equation}\label{eq:a9}
\Omega_i = W_{r\, i}\gamma_i = \sqrt{p_r^2 + m_i^2 \over 1 - v_i^2} .
\end{equation}
Because the string has no radial momentum, the quarks have equal and
opposite radial momenta.   We thus expand 
\begin{equation}
 m_2^2 - m_1^2 = \Omega_2^2(1-v_2^2) - \Omega_1^2(1-v_1^2) ,
\end{equation}
using Eq.~(\ref{eq:a1}) and Eq.~(\ref{eq:a3}), to find
\begin{equation}
\left(\Omega_2 - \Omega_1 \right)\left( M - \lambda_M \right) - \lambda_P
\left( \Omega_2 v_2 + \Omega_1 v_1\right) = m_2^2 - m_1^2.
\end{equation}
Then, we use Eqs.~(\ref{eq:a6.5}) and (\ref{eq:a7}) to obtain
\begin{eqnarray}\label{eq:a10}
&& {(M-\lambda_M)\over v_1v_2} \left[ {L\over r} - {\lambda_L\over r} +
{\lambda_P \left( v_1^2 + v_2^2 \right) \over v_1^2 - v_2^2 } \right] 
\nonumber \\
&&  \quad -\, {\lambda_P\over v_1 - v_2} \left[ {2L\over r } - {2\lambda_L
\over r} + \lambda_P{v_1 - v_2\over v_1 + v_2} \right]  \\
&&  =  {m_2^2 - m_1^2 \over v_1 -v_2} . \nonumber
\end{eqnarray}

By combining Eqs.~(\ref{eq:a8}) and (\ref{eq:a10}), we can eliminate $L/r$
in Eq.~(\ref{eq:a10}). After a bit of algebra we find
\begin{eqnarray}\label{eq:a11}
&& \left(M-\lambda_M\right)^2 + {2\lambda_P \over
v_1-v_2}\left(1-v_1v_2\right) \left(M - \lambda_M\right) + \lambda_P^2
\nonumber \\
&& = \left({m_2^2 - m_1^2 \over v_1 - v_2}\right) \left(v_1 + v_2\right) .
\end{eqnarray}
So far, we have made no approximations other than the straight string
assumption. 

\subsection{Low transverse velocity approximation}

Now we use the low velocity approximation to solve for the ratio of the
quark velocities $v_2/v_1$.  As we have seen in section \ref{sec:three},
$v_i \equiv v_{\perp\, i}$ is small if $n\gg L$, except near the inner
radial turning point. Near this point, the string angular momentum and
energy are small compared to either quark's because the string is short.
As far as the string is concerned, we can assume that $v_i \ll 1$
everywhere without changing the dynamics.  From Eqs.~(\ref{eq:lambdaP}),
(\ref{eq:lambdaL}), and (\ref{eq:lambdaM}), we may make the approximations
\begin{eqnarray}\label{eq:a12}
\lambda_P & \simeq & \frac12 ar (v_1 - v_2) , \nonumber \\
\lambda_L & \simeq & \frac13 ar^2 \left({v_1^2 - v_1v_2 + v_2^2\over v_1 +
v_2}\right) , \\
\lambda_M & \simeq & ar - {k\over r} . \nonumber
\end{eqnarray}
We note that in Eq.~(\ref{eq:a11}) we may drop the small quantities
$v_1v_2$ and $\lambda_P^2$, whereupon Eq.~(\ref{eq:a11}) reduces to
\begin{equation}\label{eq:a13}
\left(m_2^2 - m_1^2\right)  \left({v_1 + v_2\over v_1 - v_2}\right) =
\left(M-\lambda_M\right) \left( M-\lambda_M + ar \right) .
\end{equation}
We can solve Eq.~(\ref{eq:a13}) for $v_2/v_1$,
\begin{equation}\label{eq:a14}
{v_2\over v_1} = { \left(M-\lambda_M\right) \left(M-\lambda_M + ar\right) -
\left( m_2^2 - m_1^2\right) \over \left(M-\lambda_M\right)
\left(M-\lambda_M + ar\right) + \left( m_2^2 - m_1^2 \right) } .
\end{equation}
We note that if $m_1 = m_2$, then $v_2 = v_1$, as expected.  Also, if $m_2
\gg m_1$, then, since $M=m_2 + E$ and $E\ll m_2$, we find $v_2 \ll v_1$, as
one might expect.

\subsection{Solving for quark velocities}

We are now in a position to solve for the individual $v_i$'s.  We use the
small $v_i$ approximation of Eq.~(\ref{eq:a12}) in Eq.~(\ref{eq:a10}) and
again drop small terms.  These small terms are the second bracket of
Eq.~(\ref{eq:a10}), whose origin is in the $\gamma_\perp^2$ term
from Eq.~(\ref{eq:a9}). The result is
\begin{equation}\label{eq:a15}
{m_2^2 - m_1^2 \over M - \lambda_M} \> {v_1 v_2\over v_1 - v_2} = {L\over
r} + {ar\over 6}\left(v_1 +  v_2 \right).
\end{equation}
From Eq.~(\ref{eq:a14}) in the form of
\begin{eqnarray}\label{eq:a16}
{v_2\over v_1} & = & {A-B\over A+B},  \\
A & = & \left( M - \lambda_M \right) \left( M - \lambda_M + ar\right)
,\nonumber \\ 
B & = & \left( m_2^2 - m_1^2\right)\nonumber ,
\end{eqnarray}
we can substitute $v_2$ from Eq.~(\ref{eq:a16}) into Eq.~(\ref{eq:a15}) to
find
\begin{equation}\label{eq:a17}
{2L\over rv_1} = {A - B \over M - \lambda_M} - {2ar\over3}{ A \over (A+B) } .
\end{equation}
A corresponding expression for $v_2$ follows upon replacing $B$ by $-B$.

There are two important special cases of Eq.~(\ref{eq:a17}).
\begin{itemize}
\item In the the equal mass case $m_1 = m_2$ and so $B=0$. In this case
we define $E=M$ and find
\begin{equation}\label{eq:a18}
{2L\over rv_1} = E + {k\over r} - \frac23 ar  .
\end{equation}
\item In the heavy-light case, $m_2 \gg m_1$.  The total energy is $M=m_2 +
E$, where the excitation energy $E$ is small compared to $m_2$.  The
leading terms are
\begin{eqnarray}\label{eq:a19}
{2L\over rv_1}  &=&  2(E + {k\over r} - \frac23 ar)  \\
&-& {1\over m_2} \left[(E + {k \over
r})^2 - \frac53 ar(E + {k \over r}) +\frac56(ar)^2\right]. \nonumber
\end{eqnarray}
\end{itemize}

We can estimate the region of validity of the heavy-light approximation by
noting that the $1/m_2$ correction is roughly $E^2/m_2$.  We compare this
with $E$ of  the first term and conclude that if $E\ll m_2$, we may neglect
the $1/m_2$ correction.

\subsection{String corrections to quark kinetic energies}

The last step is to find the string corrections to $\Omega_i$.  The same
type of small velocity approximation used for the quarks, together with
Eq.~(\ref{eq:a17}) can be used with the $\alpha_i$ of (\ref{eq:a6.5}) and
(\ref{eq:a7}).  We find
\begin{equation}\label{eq:a20}
\alpha_1 = -\,{aL\over 3}\left({(A+3B)(M-\lambda_M)\over A^2 -B^2 -\frac23 
ar A (M-\lambda_M)}\right) .
\end{equation}
We find a similar expression for $\alpha_2$, which is obtained by replacing
$B$ by $-B$. ({\it i.e.\/}, $m_2 \leftrightarrow m_1$).
In the light equal mass limit we find 
\begin{equation}\label{eq:a21}
\alpha_1^{LL} = - \, {aL\over 3\left(E + \frac{k}r - \frac23 ar\right)}.
\end{equation}
In the heavy-light limit we obtain the same expression for $\alpha_1$ to
lowest order, and find the ${\cal O}(1/m_2)$ corrections;
\begin{eqnarray}\label{eq:a22}
\alpha_1^{HL} &= &- \, {aL\over 3\left(E + \frac{k}r - \frac23 ar\right)}
\nonumber \\
&+& {1\over m_2} {a^2 r L\over 12}{(E + {k\over r} - ar)\over (E +
{k\over r} - \frac23 ar)^2} .
\end{eqnarray}
In either case, as in the general case, if $E\gg L$
\begin{equation}\label{eq:a23}
\alpha_i {\buildrel E\gg L \over \longrightarrow} 0 ,
\end{equation}
which is the high radial excitation regime.  In this limit of $\Omega_i
\rightarrow {L/(rv_i)}$ the relativistic TCV wave equation follows.

\section*{Acknowledgment}
This work was supported in part by the US Department of Energy under
Contract No.~DE-FG02-95ER40896.

\newpage

\vspace{2cm}

\end{document}